\documentclass[%
 reprint,
 amsmath,amssymb,
 aps,
]{revtex4-1}

\usepackage{graphicx}
\usepackage{dcolumn}
\usepackage{bm}

\begin{document}

\preprint{APS/123-QED}

\title{Coupling effect of topological states and Chern insulators in two-dimensional triangular lattices}

\author{Jiayong Zhang$^{1,}$$^{2}$}
\author{Bao Zhao$^{2,}$$^{3}$}
\author{Yang Xue$^2$}
\author{Tong Zhou$^2$}
\author{Zhongqin Yang$^{2,}$$^{4,}$}
\email{zyang@fudan.edu.cn}
\address{$^1$Jiangsu Key Laboratory of Micro and Nano Heat Fluid Flow Technology and Energy Application, $\Psi_{usts}$ Institute $\&$ School of Mathematics and Physics, Suzhou University of Science and Technology, Suzhou, Jiangsu 215009, China\\
$^2$State Key Laboratory of Surface Physics and Key
Laboratory for Computational Physical Sciences (MOE) $\&$ Department
of Physics, Fudan University, Shanghai 200433, China\\
$^3$School of Physics Science and Information Technology, Shandong Key Laboratory of Optical Communication Science and Technology, Liaocheng University, Liaocheng 252059, China\\
$^4$Collaborative Innovation Center of Advanced Microstructures, Fudan University, Shanghai, 200433, China}

\date{\today}

\begin{abstract}
We investigate topological states of two-dimensional (2D) triangular lattices with multi-orbitals. Tight-binding model calculations of a 2D triangular lattice based on $\emph{p}_{x}$ and $\emph{p}_{y}$ orbitals exhibit very interesting doubly degenerate energy points at different positions ($\Gamma$ and K/K$^{\prime}$) in momentum space, with quadratic non-Dirac and linear Dirac band dispersions, respectively. Counterintuitively, the system shows a global topologically trivial rather than nontrivial state with consideration of spin-orbit coupling due to the ``destructive interference effect'' between the topological states at the $\Gamma$ and K/K$^{\prime}$ points. The topologically nontrivial state can emerge by introducing another set of triangular lattices to the system (bitriangular lattices) due to the breakdown of the interference effect. With first-principles calculations, we predict an intrinsic Chern insulating behavior (quantum anomalous Hall effect) in a family of 2D triangular lattice metal-organic framework of Co(C$_{21}$N$_{3}$H$_{15}$) (TPyB-Co) from this scheme. Our results provide a different path and theoretical guidance for the search for and design of new 2D topological quantum materials.
\end{abstract}

\pacs{Valid PACS appear here}
\maketitle





\section{\textbf{INTRODUCTION}}
The two-dimensional (2D) topological quantum states, including quantum spin Hall (QSH) states, quantum anomalous Hall (QAH) states, and their hybrid or derived states, have attracted considerable attention recently in condensed matter physics and material science due to their very special topologically protected dissipationless edge states \cite{1,2,3,4}. Up to the present, primarily two strategies have been proposed to generate the QSH state. One is the Kane-Mele model \cite{5} that first came up with a graphene system, a 2D star material with a honeycomb lattice and Dirac bands. In this model, any finite spin-orbit coupling (SOC) interaction can open a nontrivial band gap at the degenerate Dirac point and induce a QSH state. The other is the Bernevig-Hughes-Zhang model \cite{6,7} proposed first in HgTe/CdTe quantum wells, where the SOC-induced band inversion is crucial to lead to the topological state. Both of these seminal models have inspired plenty of theoretical research to find QSH states in new materials. When the time-reversal symmetry of a QSH state is broken by an appropriate internal magnetic exchange field, the QAH (Chern insulating) state is in principle expected to appear, despite actually not having the simplicity of magnetic introduction into the QSH systems \cite{8,9}. By virtue of the chiral edge states, the QSH and QAH states hold great promise in promoting the development of low-power-consumption microelectronic devices. By now, even though the QSH and QAH states have been proposed theoretically in numerous 2D material systems \cite{9,10,11,12,13,14,15,16,17,18,19,20,21,22,23,24,25,26,27,28,29,30}, the experimental observations of the QSH and QAH states have been realized only in CdTe/HgTe/CdTe or InAs/GaSb/AlSb quantum wells \cite{7,31} and Cr or V doped (Bi,Sb)$_{2}$Te$_{3}$ films \cite{8,32,33,34}, respectively. The reason is that most of the QSH and QAH effects predicted previously are very difficult to carry out in experiments under realistic conditions. Thus, finding more 2D QSH and/or QAH material systems, which are relatively facile to fabricate experimentally, and new mechanisms to realize the topological states in 2D atomic lattices beyond the hexagonal lattice is indispensable and significant.

In this work, we give a physical picture of the topological properties of a 2D triangular lattice featuring multiple \emph{p} orbitals ($\emph{p}_{x}$ and $\emph{p}_{y}$) and put forward an effective way to realize the QSH and QAH states by a tight-binding (TB) model Hamiltonian analysis. A very interesting band structure is found in the system: the two \emph{p} bands are simultaneously degenerate at the $\Gamma$ and K/K$^{\prime}$ points with quadratic non-Dirac and linear Dirac band dispersions, respectively. Previously, the Dirac bands and non-Dirac bands have been widely reported happening only separately at K/K$^{\prime}$ \cite{5,10,28} and $\Gamma$ \cite{12,27} points in different systems. More amazingly, a ``destructive interference effect'' is observed between the topological states at the $\Gamma$ and K/K$^{\prime}$ points, when the atomic SOC is taken into account in the system. The local integrated Berry curvatures around the SOC-induced gaps of $\Delta_{\Gamma}$ and $\Delta_{K/K^{\prime}}$ for the spin-up (spin-down) subspace of the TB Hamiltonian are +1 (-1) and -0.5/-0.5 (+0.5/+0.5), respectively, leading to the zero total Chern number of the system ($C$ = 0). To extract the topological behaviors at the $\Gamma$ or K/K$^{\prime}$ points, the destructive interference effect must be broken down, reached by constructing a TB-model Hamiltonian for a bitriangular lattice, proposed in this work. With first-principles calculations, we predict that the different type of 2D triangular lattice metal-organic framework Co(C$_{21}$N$_{3}$H$_{15}$) (TPyB-Co) can realize intrinsic QAH states based on this scheme. Our findings are helpful to understand deeply the topological states and provide different strategies to search for and design new QSH and QAH materials in rich 2D triangular lattice and bitriangular lattice materials.

\section{\textbf{RESULTS AND DISCUSSION}}

\subsection{\textbf{TB model for triangular lattices}}

The schematic plot of a triangular lattice with one site per unit cell is shown in Fig. 1(a). The total Hamiltonian of this triangular lattice with two orbitals ($\emph{p}_{x}$, $\emph{p}_{y}$) per lattice site can be written as
\[
\begin{array}{l}
 H(k) \\
  = H_{hop} (k) + H_{soc}  - H_M  \\
  = \left[ {\begin{array}{*{20}c}
   {H_{hop \uparrow } } & 0  \\
   0 & {H_{hop \downarrow } }  \\
\end{array}} \right] + \left[ {\begin{array}{*{20}c}
   {H_{soc \uparrow } } & 0  \\
   0 & {H_{soc \downarrow } }  \\
\end{array}} \right] \\
  - \left[ {\begin{array}{*{20}c}
   {M{\rm{\bf{I}}}} & 0  \\
   0 & { - M{\rm{\bf{I}}}}  \\
\end{array}} \right] \\
 \end{array}  \ \ \ \ \ (1)
\]
where $\emph{H}_{hop}$ expresses the nearest-neighbor (NN) hopping term of the sites, the up and down arrows indicate, respectively, the spin-up and spin-down channels, $\emph{H}_{soc}$ is the on-site SOC term, $\emph{H}_{M}$ is the magnetic exchange field term, the \emph{M} parameter is the applied exchange field strength, and \textbf{I} is a unit matrix.

The hopping terms for the spin-up and spin-down channels can both be expressed as ($\emph{H}_{hop\uparrow}=\emph{H}_{hop\downarrow}$)
\[
H_{hop \uparrow / \downarrow }  = \left[ {\begin{array}{*{20}c}
   {h_{11} } & {h_{12} }  \\
   {h_{21} } & {h_{22} }  \\
\end{array}} \right]  \ \ \ \ \ (2)
\]
in which
\[
\begin{array}{l}
 h_{11}  = \varepsilon _p  + 2t_{pp\pi } \cos (\sqrt 3 k_y ) + (\frac{3}{2}t_{pp\sigma }  \\
  + \frac{1}{2}t_{pp\pi } )\left( {\cos (\frac{3}{2}k_x  - \frac{{\sqrt 3 }}{2}k_y ) + \cos (\frac{3}{2}k_x  + \frac{{\sqrt 3 }}{2}k_y )} \right), \\
 h_{12}  = \frac{{\sqrt 3 }}{2}(t_{pp\pi }  - t_{pp\sigma } )\left( {\cos (\frac{3}{2}k_x  - \frac{{\sqrt 3 }}{2}k_y ) - } \right. \\
 \left. {\cos (\frac{3}{2}k_x  + \frac{{\sqrt 3 }}{2}k_y )} \right), \\
 h_{21}  = h_{12} , \\
 h_{22}  = \varepsilon _p  + 2t_{pp\sigma } \cos (\sqrt 3 k_y ) + (\frac{1}{2}t_{pp\sigma }  +  \\
 \frac{3}{2}t_{pp\pi } )\left( {\cos (\frac{3}{2}k_x  - \frac{{\sqrt 3 }}{2}k_y ) + \cos (\frac{3}{2}k_x  + \frac{{\sqrt 3 }}{2}k_y )} \right),(3) \\
 \end{array}
\]
In Eq. (3), $\varepsilon_{\emph{p}}$ is the on-site energy for $\emph{p}_{x}$ and $\emph{p}_{y}$ orbitals, and $t_{pp\pi}$ and $t_{pp\sigma}$ are the NN hopping parameters corresponding to the $\pi$ and $\sigma$ bonds formed by $\emph{p}_{x}$ and $\emph{p}_{y}$ orbitals.

Since the on-site SOC term does not induce the coupling between different spin components, the spin-up and spin-down spaces are separated. Thus, for the different spin subspaces, the on-site SOC terms can be written as
\[
H_{soc \uparrow / \downarrow }  = s \cdot \lambda \left[ {\begin{array}{*{20}c}
   0 & { - i}  \\
   i & 0  \\
\end{array}} \right],(4)
\]
where $\lambda$ is the atomic SOC strength, and $s=\pm 1$ represent the spin-up and spin-down components, respectively.

\begin{figure*}
\resizebox{16.5cm}{!}{\includegraphics*[93,53][697,503]{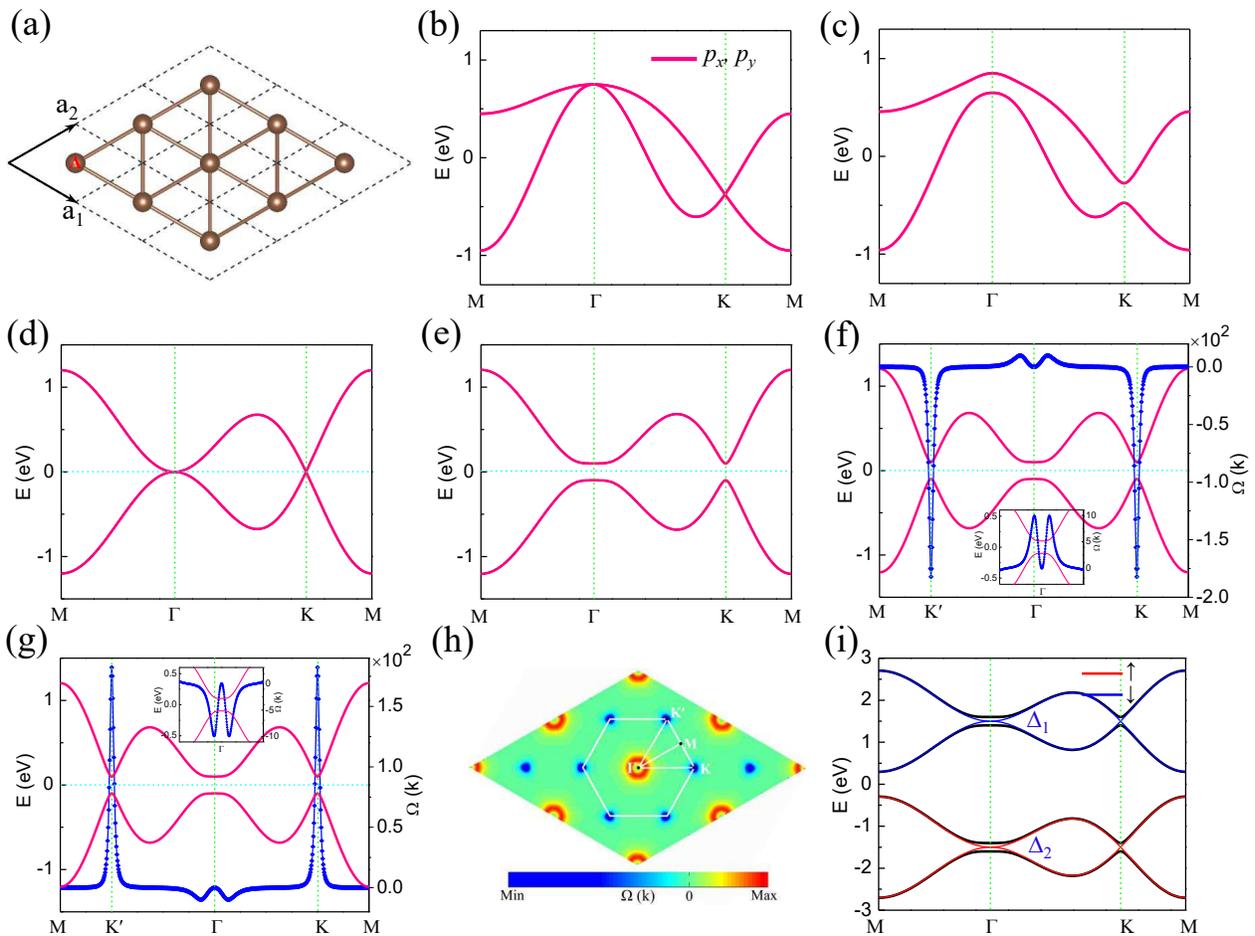}}
\caption{Bands and Berry curvatures obtained from the TB model for a triangular lattice with two orbitals ($\emph{p}_{x}$, $\emph{p}_{y}$) per lattice site. (a) Schematic plot of a triangular lattice with one site (A-site) per unit cell. The lattice unit vectors are chosen as $\vec{a_{1}}$ = $\sqrt{3}a(\sqrt{3}/2,-1/2)$ and $\vec{a_{2}}$ = $\sqrt{3}a(\sqrt{3}/2,1/2)$. (b),(c) The band structures calculated from the TB model with the parameters: $\varepsilon_{p}=0.0 eV$,$t_{pp\sigma}=0.30 eV$,$t_{pp\pi}=-0.05 eV$, and $M=0.0 eV$. $\lambda$ is (b) 0.0 eV and (c) 0.10 eV. (d),(e) The same as (b) and (c), except for $t_{pp\pi}(=-t_{pp\sigma})=-0.30 eV$. (f),(g) The Berry curvatures (blue dots) obtained for the spin-up and spin-down subspace of the TB Hamiltonian with the parameters adopted in (e), respectively. The corresponding bands are also displayed [red curves, the same as in (e)]. The insets in (f) and (g) are the magnified Berry curvatures and bands around the $\Gamma$ point in the vicinity of E$_{F}$. (h) The distribution of the Berry curvatures in 2D momentum space for the spin-up subspace of the TB Hamiltonian. (i) The TB band structure with the parameters used in (d) and (e), expect for $M=1.50 eV$. The red and blue curves denote the spin-up and spin-down bands, respectively. The black curves denote the bands with SOC ($\lambda = 0.10 eV$).}
\end{figure*}

The band structure of the triangular lattice with two orbitals ($\emph{p}_{x}$,$\emph{p}_{y}$) per lattice site can be obtained based on the above constructed TB-model Hamiltonian. The TB parameters adopted in Fig. 1(b) are $\varepsilon_{p}=0.0 eV,t_{pp\sigma}=0.30 eV,t_{pp\pi}=-0.05 eV,\lambda=0.0 eV,M=0.0 eV$, where only the NN hopping is considered. As demonstrated in Fig. 1(b), the $\emph{p}_{x}$ and $\emph{p}_{y}$ bands are not only degenerate at the $\Gamma$ point [the Brillouin zone (BZ)], with quadratic non-Dirac band dispersion, but also degenerate at the K/K$^{\prime}$ points, with linear Dirac band dispersion. These band characteristics are quite different from that obtained in the previous TB model in the honeycomb lattice\cite{5,35}. The double degeneracy at the $\Gamma$ and K/K$^{\prime}$ points is due to the C$_{6}$ symmetry owned by the triangular lattice. The different band dispersions at the $\Gamma$ and K/K$^{\prime}$ points can be comprehended with the invariant theory \cite{18,36,37}. Since the $\Gamma$ point is a time reversal invariant point, the linear k term in the energy eigenvalue at this point is forbidden, while there is no such limitation at the K and K$^{\prime}$ points [37]. After the on-site SOC is included in the calculations, the degeneracy of the quadratic non-Dirac bands at the $\Gamma$ point and linear Dirac bands at the K/K$^{\prime}$ points is lifted and band gaps are opened around these points by any finite value of $\lambda $, as illustrated in Fig. 1(c) with $\lambda=0.1 eV$. If we set the hopping parameters $t_{pp\pi}=-t_{pp\sigma}$, the degenerate bands at the $\Gamma$ point and K/K$^{\prime}$ points without SOC are located at the same energy position [see Fig. 1(d)]. After the atomic SOC is taken into account, a desirable global band gap is opened, as shown in Fig. 1(e) (with $\lambda=0.1 eV$).

\subsection{\textbf{Coupling effect of topological states}}

We now focus on the topological behaviors of the system. Berry curvatures and Chern numbers are calculated to identify the topological properties of the global SOC-induced band gap in the 2D triangular lattice with the two $\emph{p}_{x}$ and $\emph{p}_{y}$ orbitals per lattice site [Fig. 1(e)]. The Berry curvature $\Omega (\mathbf{k})$ is calculated by \cite{38,39}
\begin{equation*}
\Omega (\mathbf{k})=\sum\limits_{n}{{f_{n}}}{\Omega _{n}}(\mathbf{k}),
\end{equation*}%
\begin{eqnarray*}
\Omega _{n}(\mathbf{k}) &=&-2\mathrm{Im}\sum_{m\neq n}\frac{\hbar ^{2}\left\langle
\psi _{n\mathbf{k}}|v_{x}|\psi _{m\mathbf{k}}\right\rangle \left\langle \psi
_{m\mathbf{k}}|v_{y}|\psi _{n\mathbf{k}}\right\rangle }{%
(E_{m}-E_{n})^{2}},\ \ \ \ \ \ (5)
\end{eqnarray*}%
where $E_{n}$ is the eigenvalue of${\left\vert {{\psi_{n\mathbf{k}}}}%
\right\rangle}$, $f_{n}$ is the Fermi-Dirac distribution function, $\upsilon _{x}$ and $\upsilon _{y}$ are the velocity operators, and the summation is over all of the occupied states. The Chern number $C$ is obtained by integrating the $\Omega (\mathbf{k})$ over the first BZ, $C = \frac{1}{{2\pi }}\sum\limits_{n}{\int_{BZ}{{d^{2}}k}}{\Omega _{n}}$. Since the spin-up and spin-down subspaces are decoupled, one can calculate the Berry curvatures separately for the spin-up and spin-down subspaces of the TB Hamiltonian.

Figure 1(f) displays the calculated Berry curvatures (blue dots) for the spin-up subspace Hamiltonian with the parameters used in Fig. 1(e). The corresponding bands are also shown (red curves). Clearly, the nonzero Berry curvatures distribute around all the SOC-induced gaps at the $\Gamma$, K, and K$^{\prime}$ points. In this work, the values of the Berry curvatures integrated around the $\Gamma$, K, and K$^{\prime}$ points are labeled as the $C_{\Gamma}$, $C_{K}$, and $C_{K^{\prime}}$, respectively. By integrating the $\Omega (\mathbf{k})$ around the $\Gamma$, K, and K$^{\prime}$ points, the nonzero numbers of $C_{\Gamma\uparrow}=1$, $C_{K\uparrow}=-0.5$, and $C_{K^{\prime}\uparrow}=-0.5$ are obtained, respectively. Obviously, the Berry curvatures around the K and K$^{\prime}$ points have the same sign and both are negative, while the $\Omega (\mathbf{k})$ around the $\Gamma$ point are positive [see the inset of Fig. 1(f)] and have opposite sign to those of the K and K$^{\prime}$ points. The opposite distribution tendency around the $\Gamma$ and K/K$^{\prime}$ points can be seen more explicitly in the corresponding distribution of $\Omega (\mathbf{k})$ in the 2D momentum space in Fig. 1(h). The $\Omega (\mathbf{k})$ distribution in the whole BZ [Fig. 1(h)] also emerges the C$_{6}$ symmetry, owned by the triangular lattice [Fig. 1(a)]. From the obtained $C_{\Gamma\uparrow}=1$, $C_{K\uparrow}=-0.5$, and $C_{K^{\prime}\uparrow}=-0.5$, we have $C_{non-Dirac\uparrow}=C_{\Gamma\uparrow}=1$ and $C_{Dirac\uparrow}=C_{K\uparrow}+C_{K^{\prime}\uparrow}=-1$. Due to the opposite signs of Chern numbers of the non-Dirac and Dirac bands, the total Chern number for the spin-up subspace in the system is zero, i.e., $C_{\uparrow}=C_{non-Diarc\uparrow}+C_{Diarc\uparrow}=0$.

Similarly, Fig. 1(g) shows the corresponding results for the spin-down subspace Hamiltonian, in which the Berry curvatures around the $\Gamma$, K, and K$^{\prime}$ points are all opposite to the case of the spin-up subspace; namely, $C_{\Gamma\uparrow}=-1$, $C_{K\downarrow}=+0.5$, and $C_{K^{\prime}\downarrow}=+0.5$. The total Chern number for the spin-down subspace is also $C_{\downarrow}=C_{non-Dirac\downarrow}+C_{Dirac\downarrow}=C_{\Gamma\downarrow}+C_{K\downarrow}+C_{K^{\prime}\downarrow}=0$. Thus, the spin Chern number, defined as $C_{s}=\frac{1}{2}(C_{\uparrow}-C_{\downarrow})$, is $C_{s}$=0, indicating that the global SOC-induced gap in the 2D triangular lattice [Fig. 1(e)] is a topologically trivial state.

However, for the non-Dirac band around the $\Gamma$ point, the spin Chern number $C_{non-Dirac,s}=\frac{1}{2}(C_{non-Dirac\uparrow}-C_{non-Diarc\downarrow})=\frac{1}{2}(C_{\Gamma\uparrow}-C_{\Gamma\downarrow})=1$. And for the Dirac bands around the K and K$^{\prime}$ points, the spin Chern number $C_{Dirac,s}=\frac{1}{2}(C_{Diarc\uparrow}-C_{Diarc\downarrow})=\frac{1}{2}((C_{K\uparrow}+C_{K^{\prime}\uparrow})-(C_{K\downarrow}+C_{K^{\prime}\downarrow}))=-1$. Thus, the local band gaps opened for the non-Dirac band and Dirac bands around the $\Gamma$ and K/K$^{\prime}$ points are both topologically nontrivial and can present the QSH effect if they can be measured individually in experiments. The nontrivial non-Dirac band gap around the $\Gamma$ point can provide a chiral spin current flowing along the sample edges. They are also so for the Dirac band gaps around the K and K$^{\prime}$ points. Since the spin Chern numbers for these two types of band gaps are opposite ($C_{non-Diarc,s}=1$,$C_{Diarc,s}=-1$), the chiralities of the spin currents provided by the two types of band gaps are opposite. Thus, the two sets of the topology-protected chiral edge states cancel each other and a global topologically trivial state is finally acquired for the whole system. The global topologically trivial state of the above system is the result of the coupling effect, dubbed as destructive interference effect, between the two topologically nontrivial states separately around the K/K$^{\prime}$ points with Dirac electrons and around the $\Gamma$ point with non-Dirac electrons due to the opposite Berry curvatures. Similarly, the global topological property of a system may be enhanced, with a high Chern number, when a ``constructive interference effect'' happens. The proposed interference effect expression is a vivid analogy to that of light waves and helpful to intuitively understand the global topological behavior of a system when there are two or more nontrivial states around the E$_{F}$ in the system. Here we present the coupling effect of topological states; usually attention to the topology of the bands is merely around one point (such as at $\Gamma$ \cite{12,27}) or one pair of points (such as K and K$^{\prime}$ \cite{5,10,28}) in the first BZ. The concept of the coupling effect pushes the understanding of the topological states one big step ahead, and has a great impact on the global topological behavior if two or more nontrivial gaps opened around the E$_{F}$ in the systems.

When a relatively large exchange field is considered in the model, the spin-up and spin-down bands can be completely separated from each other, as shown in Fig. 1(i) with $M = 1.5 eV$. Similarly, due to the destructive interference effect, the Chern numbers of the two separated SOC-induced gaps for the whole spin-up and spin-down bands are $C_{\uparrow}=0$ and $C_{\downarrow}=0$. Hence, the two band gaps opened at about 1.5 and -1.5 eV are both topologically trivial states rather than Chern insulating states. Besides the \emph{p} orbitals ($\emph{p}_{x}$ and $\emph{p}_{y}$), the interference effect can also happen in a 2D triangular lattice with two \emph{d} orbitals per lattice site, such as ($\emph{d}_{xy}$,$\emph{d}_{x^{2}-y^{2}}$) or ($\emph{d}_{xz}$ and $\emph{d}_{yz}$) orbitals, accommodated on the triangular lattice with C$_{6}$ symmetry. Thus, we provide very clear physical pictures of the topological properties of the SOC-induced gap in 2D triangular lattices with two \emph{p} or \emph{d} orbitals.

\subsection{\textbf{Topological states in bitriangular lattices}}

\begin{figure*}
\resizebox{16.5cm}{!}{\includegraphics*[90,51][697,504]{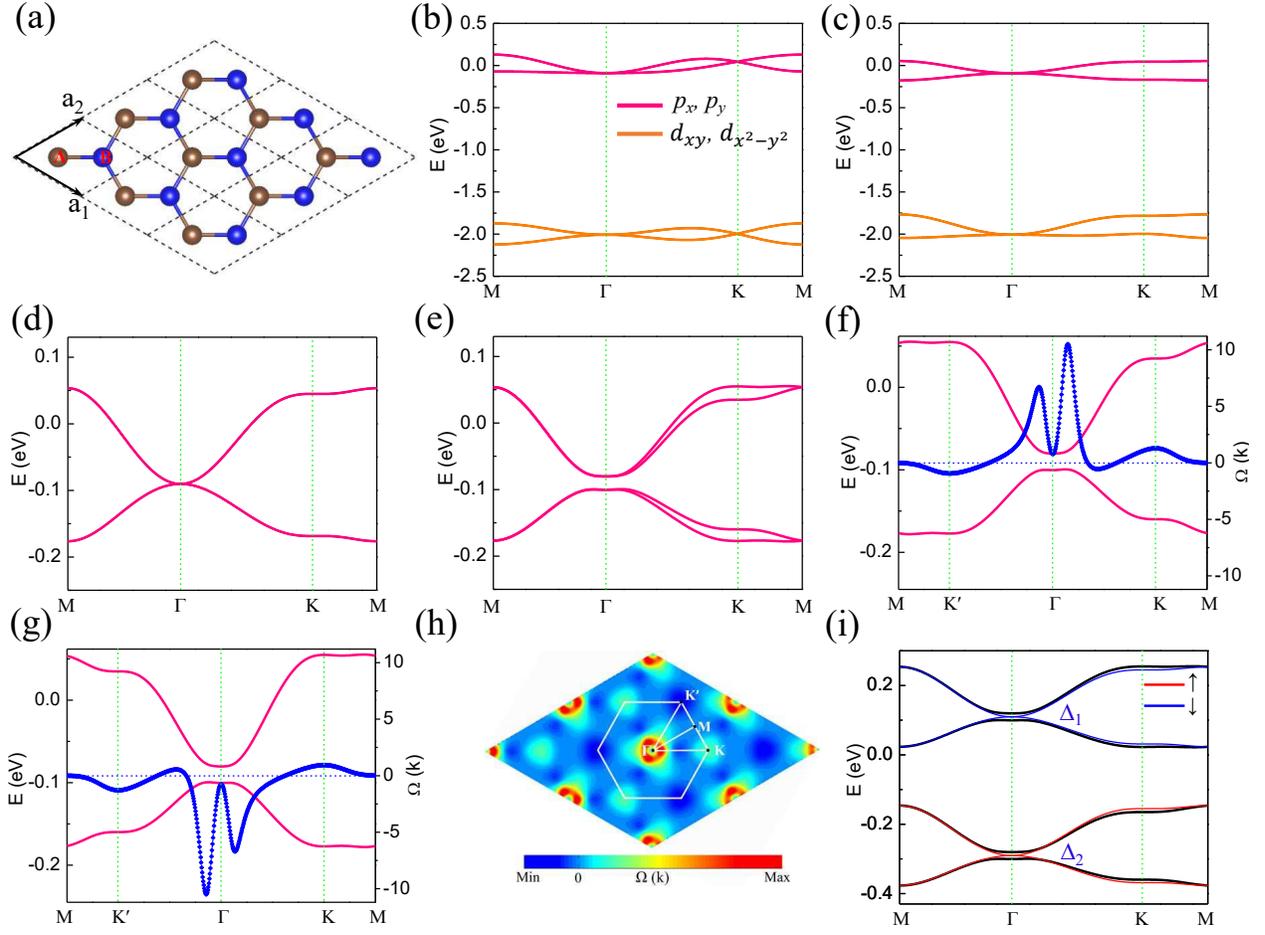}}
\caption{Bands and Berry curvatures obtained from the TB model for the bitriangular lattice formed by two triangular lattices with different orbitals. One is with two orbitals of $\emph{p}_{x}$ and $\emph{p}_{y}$ per lattice site and the other is with two orbitals of $\emph{d}_{xy}$ and $\emph{d}_{x^{2}-y^{2}}$ per lattice site. (a) Schematic plot of the bitriangular lattice with two nonequivalent sites (A-site and B-site) per unit cell. The lattice unit vectors are chosen as $\vec{a_{1}}$ = $\sqrt{3}a(\sqrt{3}/2,-1/2)$ and $\vec{a_{2}}$ = $\sqrt{3}a(\sqrt{3}/2,1/2)$. (b) The band structure calculated from the TB model with the parameters: $\varepsilon{_p} = 0.0 eV, t_{pp\sigma} = -0.04 eV, t_{pp\pi} = 0.01 eV, M_{A}=0.0 eV, \lambda_{A}=0.0 eV, \varepsilon_{d} = -2.00 eV,t_{dd\delta} = -0.01 eV, t_{dd\sigma} = -0.04 eV, t_{dd\pi} = 0.03 eV, M_{B} = 0.0 eV, \lambda_{B} = 0.0 eV, t_{pd\sigma} = 0.0 eV, t_{pd\pi} = 0.0 eV$. (c) The same as (b), except for $t_{pd\sigma} = 0.25 eV$  and $t_{pd\pi} = 0.20 eV$. (d) The magnified bands near the E$_{F}$ in (c), formed by the $\emph{p}_{x}$ and $\emph{p}_{y}$ orbitals. (e) The same as (d), except for $\lambda_{A} = 0.01 eV$ and $\lambda_{B}=0.01 eV$. (f),(g) The band structures (pink curves) and Berry curvatures (blue dots) obtained for the spin-up and spin-down subspaces of the TB Hamiltonian with the parameters adopted in (e), respectively. The bands are magnified near the band gap induced by the SOC interaction. In (b)--(g), the pink and orange colors indicate the bands contributed by $\emph{p}$ and $\emph{d}$ orbitals, respectively. (h) The distribution of the Berry curvatures in 2D momentum space for the spin-up subspace of the TB Hamiltonian. (i) The TB bands with the parameters used in (d) and (e), expect for $M_{A} = 0.20 eV$, $M_{B} = 0.20 eV$. The red and blue curves denote the spin-up and spin-down bands, respectively. The black curves denote the bands with SOC ($\lambda_{A} = 0.01 eV$ and $\lambda_{B} = 0.01 eV$).}
\end{figure*}

To acquire topologically nontrivial states in the 2D triangular lattice with $\emph{p}_{x}$ and $\emph{p}_{y}$ orbitals, one direct tactic is eliminating the destructive interference effect through destroying the non-Dirac or Dirac band degeneracy, respectively, at the $\Gamma$ and K/K$^{\prime}$ points without SOC. Since the symmetry of the BZ center ($\Gamma$ point) is higher than that of the K and K$^{\prime}$ points, it is relatively easy to break the Dirac band degeneracy at the K and K$^{\prime}$ points. Therefore, if the band gaps opened at the K and K$^{\prime}$ points become topological trivial, with $C_{K\uparrow}+C_{K^{\prime\uparrow}}=C_{K\downarrow}+C_{K^{\prime\downarrow}}=0$, the whole Chern numbers for the spin-up and spin-down subspaces are then determined by the corresponding non-Dirac band around the $\Gamma$ point: $C_{\uparrow}=C_{\Gamma\uparrow}+C_{K\uparrow}+C_{K^{\prime}\uparrow}=C_{\Gamma\uparrow}=+1$ and $C_{\downarrow}=C_{\Gamma\downarrow}+C_{K\downarrow}+C_{K^{\prime}\downarrow}=C_{\Gamma\downarrow}=-1$. For the whole system, $C_{s}=\frac{1}{2}(C_{\uparrow}-C_{\downarrow})=1$, indicating the appearance of the topologically nontrivial phase.

A different TB-model Hamiltonian for a bitriangular lattice formed by two triangular lattices ($\emph{p}$-$\emph{d}$) is constructed to destroy the Dirac bands around the K and K$^{\prime}$ points. Figure 2(a) shows a schematic plot of the bitriangular lattice formed by two different (A and B) triangular lattices. The unit cell of this lattice contains two nonequivalent sites (A-site and B-site). The A-site and B-site in the ``bitriangular lattice'' are occupied by different atoms. Thus, the atomic orbitals and on-site energies of the A-site and B-site are different, while in the previous widely studied ``honeycomb lattice''\cite{5,26,27,28,35}, same atoms with the same orbital types and parameters are localized at the A-site and B-site. Therefore, the bitriangular lattice studied in this work is very different from the traditional honeycomb lattice. The TB model is built with the consideration of two orbitals of ($\emph{p}_{x}$, $\emph{p}_{y}$) on each A-site and two orbitals of ($\emph{d}_{xy}$,$\emph{d}_{x^{2}-y^{2}}$) on each B-site. In this bitriangular lattice, the NN hopping happens between the nonequivalent A and B triangular lattices, while the next-nearest-neighbor (NNN) hopping occurs within the A or B triangular lattice. Similar to the simple 2D triangular lattice above, the total Hamiltonian of the bitriangular lattice in the basis of ($\emph{p}_{x}$, $\emph{p}_{y}$, $\emph{d}_{xy}$, $\emph{d}_{x^{2}-y^{2}}$) can be written as
\[
\begin{array}{l}
 H_{bi} (k) = H_{bi,hop} (k) + H_{bi,soc}  - H_{bi,M}  \\
  = \left[ {\begin{array}{*{20}c}
   {H_{bi,hop \uparrow } } & 0  \\
   0 & {H_{bi,hop \downarrow } }  \\
\end{array}} \right] + \left[ {\begin{array}{*{20}c}
   {H_{bi,soc \uparrow } } & 0  \\
   0 & {H_{bi,soc \downarrow } }  \\
\end{array}} \right] \\
  - \left[ {\begin{array}{*{20}c}
   {\begin{array}{*{20}c}
   {M_A {\rm{I}}} & 0  \\
   0 & {M_B {\rm{I}}}  \\
\end{array}} & 0  \\
   0 & {\begin{array}{*{20}c}
   { - M_A {\rm{I}}} & 0  \\
   0 & { - M_B {\rm{I}}}  \\
\end{array}}  \\
\end{array}} \right],(6) \\
 \end{array}
\]
where $H_{bi,hop}$ expresses the hopping term (including the NN and NNN interactions) of the bitriangular lattice, $H_{bi,soc}$ is the on-site SOC term, $H_{bi,M}$ is the magnetic exchange field term in the lattice, and $M_{A}$ and $M_{B}$ are the applied exchange field strengths on the A-site and B-site, respectively. The concrete expression of the hopping term of the bitriangular lattice is given in the Supporting Note 1 of Supplemental Material \cite{40}. More parameters are needed to describe the hopping term in the bitriangular lattice, including $\varepsilon_{p}$ and $\varepsilon_{d}$ (the onsite energies for $\emph{p}_{x}$/$\emph{p}_{y}$ and $\emph{d}_{xy}$/$\emph{d}_{x^{2}-y^{2}}$ orbitals, respectively), $t_{pp\sigma}$, $t_{pp\pi}$, $t_{dd\delta}$, $t_{dd\sigma}$, and $t_{dd\pi}$ (the NNN hopping parameters), and $t_{pd\sigma}$ and $t_{pd\pi}$  (the NN hopping parameters).

For the different spin subspaces, the on-site SOC is written as a $4 \times 4$ matrix:
\[
H_{bi,soc \uparrow / \downarrow }  = s \cdot \left[ {\begin{array}{*{20}c}
   0 & { - \lambda _A i} & 0 & 0  \\
   {\lambda _A i} & 0 & 0 & 0  \\
   0 & 0 & 0 & {2\lambda _B i}  \\
   0 & 0 & { - 2\lambda _B i} & 0  \\
\end{array}} \right],(7)
\]
where $\lambda_{A}$ and $\lambda_{B}$ is the atomic SOC strength of the A-site and B-site, respectively. $s=\pm1$ represent the spin-up and spin-down components, respectively. The spin-up and spin-down spaces are also decoupled as discussed for the simple triangular lattice.

The band structures of the bitriangular lattice given by above the TB Hamiltonian are given in Fig. 2, where the $\emph{p}_{x}$/$\emph{p}_{y}$ bands are considered not to mix with the $\emph{d}_{xy}$/$\emph{d}_{x^{2}-y^{2}}$ bands. We first neglect the NN hopping and only consider the NNN hopping, with which the band structure is formed by the bands of two irrelevant triangular lattices, as shown in Fig. 2(b). The TB parameters set in Fig. 2(b) are $\varepsilon_{p}=0.0 eV$, $t_{pp\sigma}=-0.04 eV$, $t_{pp\pi}=0.01 eV$, $M_{A}=0.0 eV$, $\lambda_{A}=0.0 eV$, $\varepsilon_{d}=-2.00 eV$,$t_{dd\delta}=-0.01 eV$, $t_{dd\sigma}=-0.04 eV$, $t_{dd\pi}=0.03 eV$,  $M_{B}=0.0 eV$, $\lambda_{B}=0.0 eV$, $t_{pd\sigma}=0.0 eV$, $t_{pd\pi}=0.0 eV$. Note that both of the NN hoping parameters of $t_{pd\sigma}$ and $t_{pd\pi}$ are chosen to be zero, indicating no interaction between the A-B sublattices. The pink and orange colors in Fig. 2(b) give the bands formed by the $\emph{p}$ and $\emph{d}$ orbitals, respectively. Similar to the bands in Fig. 1(b), both of the $\emph{p}$ and $\emph{d}$ bands in Fig. 2(b) are not only degenerate at the $\Gamma$ point with quadratic non-Dirac band dispersion but also degenerate at the K/K$^{\prime}$ points with linear Dirac band dispersion. When the hopping between the two nonequivalent triangular lattices (the NN hopping) is turned on, the degeneracy at the $\Gamma$ point still keeps while the degeneracy at the K and K$^{\prime}$ points is eliminated, with large band gaps opened at the K and K$^{\prime}$ points, as displayed in Fig. 2(c) by setting $t_{pd\sigma}=0.25 eV$ and  $t_{pd\pi}=0.20 eV$ (The bands at the K$^{\prime}$ point are not shown). The above band feature can be seen more clearly in Fig. 2(d), displaying the magnified bands near the E$_{F}$ of Fig. 2(c), coming from the triangular lattice with $\emph{p}_{x}$ and $\emph{p}_{y}$ orbitals. Therefore, the NN hopping between the two nonequivalent A-B triangular lattices in the bitriangular lattice can destroy the Dirac bands and open large nontrivial band gaps at the K and K$^{\prime}$ points. When the on-site SOC term is applied, the quadratic non-Dirac band touching at the $\Gamma$ point is removed and a band gap is opened [see Fig. 2(e), with $\lambda_{A}=0.01 eV$ and $\lambda_{B}=0.01 eV$]. As expected, the destructive interference effect of the topological states is successfully eliminated in the bitriangular lattice, resulting in the whole topology of the system determined solely by the SOC-induced gap around the $\Gamma$ point, as discussed below.

Figure 2(f) shows the calculated Berry curvatures (blue dots) for the spin-up subspace Hamiltonian with the parameters adopted in Fig. 2(e). The $\Omega (\mathbf{k})$ primarily distributes around the SOC-induced gap at the $\Gamma$ point with the similar distribution feature around the $\Gamma$ point in Fig. 1(f), while the $\Omega (\mathbf{k})$ distribution around the K and K$^{\prime}$ points are now opposite with small values. The distribution of $\Omega (\mathbf{k})$ in 2D momentum space is given in Fig. 2(h), where the $\Omega (\mathbf{k})$ around the $\Gamma$ point exhibits an approximate ring pattern and the $\Omega (\mathbf{k})$ around the K and K$^{\prime}$ points are very small and with opposite signs. Different from the C$_{6}$ symmetry owned in Fig. 1(h), the $\Omega (\mathbf{k})$ distribution in Fig. 2(h) satisfies C$_{3}$ symmetry, consistent with the symmetry of the bitriangular lattice [Fig. 2(a)]. Actually, it is the absence of the inversion symmetry of the bitriangular lattice that causes the large band gaps opened at the K and K$^{\prime}$ points [Fig. 2(c) and 2(d)] and opposite Berry curvature values around these two points [Fig. 2(f) or Fig. 2(g)]. Integrating the $\Omega (\mathbf{k})$ around the $\Gamma$ and K/K$^{\prime}$ points, the Chern numbers of  $C_{\Gamma\uparrow}=+1$ and $C_{K\uparrow}+C_{K^{\prime}\uparrow}=0$ are obtained, and the total Chern number for the spin-up subspace is $C_{\uparrow}=C_{\Gamma\uparrow}+C_{K\uparrow}+C_{K^{\prime}\uparrow}=C_{\Gamma\uparrow}=+1$. Figure 2(g) presents the $\Omega (\mathbf{k})$ for the spin-down subspace (blue dotted curve), and the total Chern number for the spin-down subspace is $C_{\downarrow}=C_{\Gamma\downarrow}+C_{K\downarrow}+C_{K^{\prime}\downarrow}=C_{\Gamma\downarrow}=-1$. Therefore, the spin Chern number is $C_{s}=\frac{1}{2}(C_{\uparrow}-C_{\downarrow})=1$, indicating the QSH state possessed by this bitriangular lattice. The acquisition of the topologically nontrivial state in this lattice can be ascribed to the breakdown of the destructive interference effect since the topological state at the K/K$^{\prime}$ points is destroyed by the $\emph{p}$-$\emph{d}$ NN interaction between the two sets of the triangular lattices [Figs. 2(c) and 2(d)]. When a magnetic exchange field $\emph{M}$ is considered in the TB model, the spin-up and spin-down bands split [see Fig. 2(i), with $M_{A}=0.20 eV$, $M_{B}=0.20 eV$]. Clearly, the Chern numbers of $C=+1$ and $C=-1$ can be achieved at the spin-up and spin-down band gaps, respectively. And Chern insulators will be obtained when the E$_{F}$ is located inside these SOC-induced band gaps ($\Delta_{1}$ and $\Delta_{2}$).

\begin{figure*}
\resizebox{16.5cm}{!}{\includegraphics*[79,128][794,503]{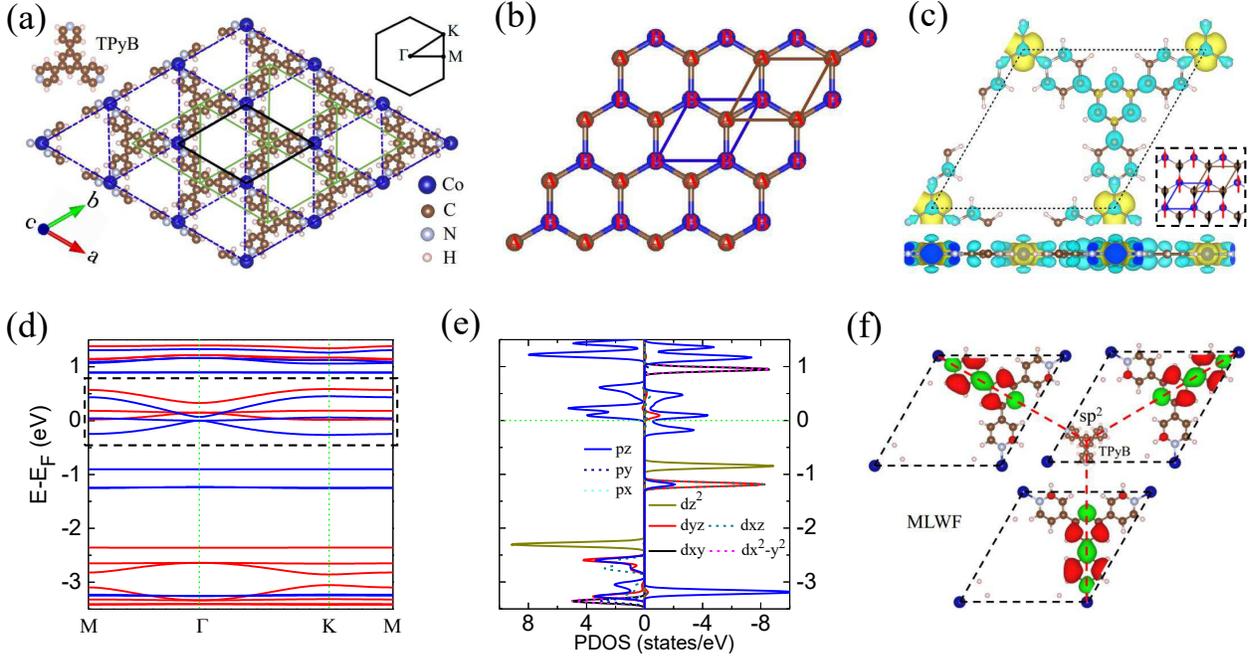}}
\caption{Crystal structures and electronic properties of TPyB-Co. (a) The atomic structure of the TPyB-Co lattice. The top right inset shows the first Brillouin zone and the special k points with high symmetries. The black solid lines indicate the unit cell. The blue dashed and green solid lines outline the triangular lattices formed by the Co atoms and TPyB molecules, respectively. (b) A schematic drawing depicting the bitriangular lattice formed by A and B sublattices. The A and B sublattices represent the lattices formed by TPyB molecules and Co atoms, respectively. (c) The spatial distribution of spin-polarized electron density for the FiM ground state of TPyB-Co. The yellow and light blue colors give the net spin-up and spin-down charge densities, respectively. The inset denotes the spin directions of the AB-sublattices in the bitriangular lattice. (d) Band structure of TPyB-Co without SOC. The red and blue curves denote the spin-up and spin-down bands, respectively. (e) The PDOSs of TPyB-Co. The positive and negative values of PDOSs express the spin-up and spin-down components, respectively. (f) The calculated three MLWFs corresponding to the three DFT spin-up bands in the rectangular dashed box in (d).}
\end{figure*}

To remove the interference effect of the topological states, besides the ($\emph{d}_{xy}$, $\emph{d}_{x^{2}-y^{2}}$) orbitals, the B-site can also own other orbitals, such as ($\emph{p}_{x}$, $\emph{p}_{y}$) or $\emph{d}_{z^{2}}$ orbitals. Due to the existence of the NN hopping between the two triangular lattices, the topological states at the K/K$^{\prime}$ points can be broken and the interference effect collapses down. Thus, the topological behaviors of the whole system are merely determined by the topological properties around the $\Gamma$ point. Note that if the B-site is set to own the ($\emph{d}_{xz}$, $\emph{d}_{yz}$) orbitals, the destructive interference effect cannot be eliminated due to the zero NN hopping between ($\emph{p}_{x}$, $\emph{p}_{y}$) and ($\emph{d}_{xz}$, $\emph{d}_{yz}$) orbitals in the 2D bitriangular lattice. The calculated TB band structures of the 2D bitriangular lattice with the ($\emph{d}_{xz}$, $\emph{d}_{yz}$) or $\emph{d}_{z^{2}}$ orbitals at the B-site are given in the Supplemental Material \cite{40}. Hence, the underlying physics for the additional triangular lattice in the bitriangular lattice to lead to the QSH or QAH states is whether the hopping interaction from this additional lattice can open a trivial band gap at the K/K$^{\prime}$ points before the SOC is considered. Thus, appropriate $\emph{p}$ or $\emph{d}$ orbitals must be chosen at the B-site in the bitriangular lattice to remove the interference effect and make the whole system emerge topological behaviors.

\subsection{\textbf{Electronic structures and topological properties of TPyB-Co}}

The geometry optimization and electronic structure calculations of the TPyB-Co are performed with projected augmented wave (PAW) \cite{41} formalism by using the first-principles calculations method based on density functional theory, as implemented in the Vienna \emph{ab-initio} simulation package (VASP) \cite{42}. The Perdew-Burke-Ernzerhof generalized-gradient approximation (GGA-PBE) is employed to describe the exchange and correlation functional \cite{43}. The GGA+U method \cite{44} is adopted to take into account the correlation effects of the Co 3$\emph{d}$ electrons, and the on-site Coulomb interaction $\emph{U}$ and the exchange interaction $\emph{J}$ are set to be 4.0 and 0.9 eV, respectively. All calculations are performed with a plane-wave cutoff energy of 450 eV and a vacuum space of larger than 15 {\AA} is used to avoid the interaction between neighboring slabs. The convergence criterion for the total energy is set to be 10$^{-6}$ eV and all the atoms in the TPyB-Co unit cell are allowed to relax until the Hellmann-Feynman force on each atom is smaller than 0.01 eV/{\AA}. In addition, a sufficiently dense Monkhorst-Pack k-point grid of $10 \times 10 \times 1$ is adopted. The Berry curvatures of TPyB-Co are calculated on the basis of Wannier functions. We first construct the maximally localized Wannier functions (MLWFs) \cite{45,46} through a non-self-consistent calculations on a $6 \times 6 \times 1$ k-point grid based on the previously converged self-consistent DFT charge potential, as implemented in the Wannier90 package \cite{47}. The numerical algorithm described in Ref. \cite{48} is then employed to calculate the Berry curvatures.

\begin{figure}
\resizebox{8.5cm}{!}{\includegraphics*[89,98][573,518]{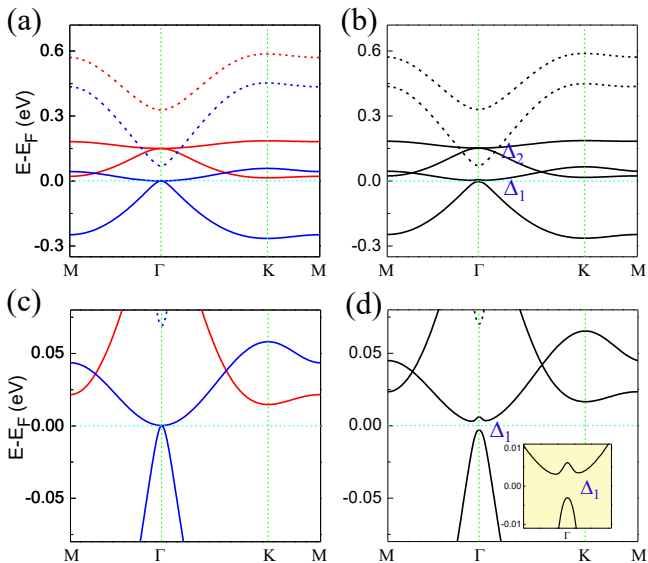}}
\caption{Featured band structures of TPyB-Co. (a) Zoom of the band structure in the rectangular dashed box in Fig. 3d. The red and blue curves denote the spin-up and spin-down bands, respectively. The dotted curves express the bands formed by the \emph{s} molecular orbitals of TPyB. (b) The corresponding band structure of (a) by including the SOC. (c),(d) Magnification of the bands near the E$_{F}$ in (a) and (b), respectively. The inset in (d) is the magnified bands around the gap induced by the SOC interaction.}
\end{figure}

\begin{figure}
\resizebox{8.5cm}{!}{\includegraphics*[48,200][336,539]{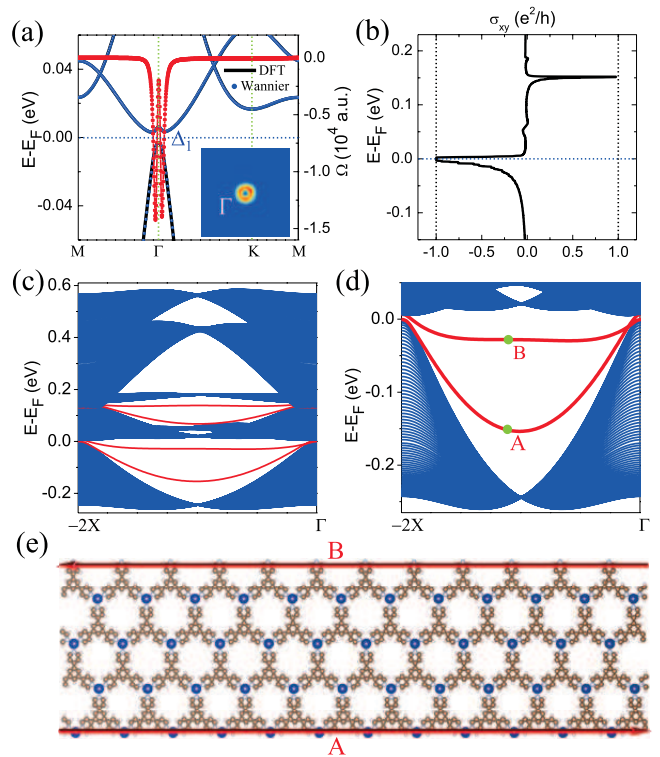}}
\caption{Intrinsic Chern insulating state for TPyB-Co. (a) The calculated band structure for TPyB-Co, obtained with DFT methods (black solid curves) and Wannier interpolation approaches (blue dots). The SOC interaction is considered. The red dots denote the Berry curvatures of TPyB-Co. The inset presents the distribution of the Berry curvatures around the $\Gamma$ point in the 2D Brillouin zone. (b) The calculated anomalous Hall conductivity $\sigma_{xy}$ as a function of the energy. (c) The calculated bands of a TPyB-Co ribbon system based on the Wannierized Hamiltonian. (d) The magnified bands near the E$_{F}$ in (c). The blue curves denote the projected bulk energy bands, while the red curves (A and B) represent the chiral edge states localized on the two opposite sides of the ribbon. (e) A schematic drawing depicting the TPyB-Co ribbon. The red arrows on the opposite sides of the ribbon represent the edge states A and B, respectively.}
\end{figure}

We now systematically investigate the electronic structures and topological properties of a 2D metal-organic framework with the bitriangular lattice from density functional theory (DFT) calculations and justify that topologically nontrivial states can be realized in realistic material systems with the above proposed mechanism. The topologically nontrivial states have been proposed in some 2D metal-organic frameworks \cite{26,27,28,29,30}, with certain special lattice structures such as the honeycomb \cite{26,27,28} and kagome lattices \cite{29,30}, formed by metal atoms or organic molecules. The topological states in the 2D honeycomb lattice metal-organic\cite{26,27,28} frameworks come from the linear Dirac bands at the K/K$^{\prime}$ points. Here, we explore a type of 2D triangular lattice metal-organic framework Co(C$_{21}$N$_{3}$H$_{15}$) (TPyB-Co), whose unit cell contains one TPyB (C$_{21}$N$_{3}$H$_{15}$) molecule and one Co atom, as shown in Fig. 3a. The topological states in this type of 2D triangular lattice metal-organic frameworks (TPyB-Co) come from the combination effect between Dirac and non-Dirac bands (as shown in the above constructed TB models), totally different from the previous widely studied 2D honeycomb lattices \cite{26,27,28}. The TPyB-Co can be treated as the bitriangular lattice discussed above [Fig. 3(b)], where the TPyB molecule can be regarded as the A-site and the Co atom can be regarded as the B-site. Therefore, the TPyB and Co form two nonequivalent triangular lattices. The atomic structure of TPyB-Co is the same as Fe(C$_{21}$N$_{3}$H$_{15}$) (TPyB-Fe) which has been successfully synthesized in recent experiments \cite{49}. The structural stability of the TPyB-Co is examined by a first-principles molecular dynamics simulation, as displayed in Fig. S1 of the Supplemental Material [40]. The optimized lattice constant in the \emph{ab} plane is $a=13.48 {\AA}$ and the N-Co distance is about 1.996 {\AA}, indicating the strong chemical binding between the Co atom and the TPyB molecule. Interestingly, due to the interaction between the Co atom and TPyB molecule, the two 4\emph{s} electrons of the Co atom first transfer to its 3$\emph{d}$ orbital, then one electron of Co atom transfers to the molecular orbital of TPyB. The five spin-up occupied \emph{d} states and three spin-down occupied $\emph{d}_{xz}$, $\emph{d}_{yz}$, and $\emph{d}_{z^{2}}$ states give rise to a 2.0 $\mu_{B}$ local magnetic moment for a Co atom in the system. In addition, the spin-down \emph{p} orbitals just below the E$_{F}$ [Fig. 3(e)], occupied by one electron transferred from the Co atom, lead to a -1.0 $\mu_{B}$ local magnetic moment for the C and N atoms, which together with the magnetic moment of the Co atom results in a 1.0 $\mu_{B}$ net magnetic moment in the system. The total energy calculations show that the magnetic ground state of the system is ferrimagnetic (FiM), with the local magnetic moments of Co and TPyB aligning in an anti-parallel style, as indicated in Fig. 3(c). This FiM state is more stable than the ferromagnetic (FM) and nonmagnetic states by about 28.8 meV/unit cell and 1.92 eV/unit cell, respectively. The spatial distribution of the FM spin-polarized electron density is given in Fig. S2(a) of the Supplemental Material. In the following, the electronic properties of the FiM TPyB-Co are chiefly investigated.

The calculated band structures and atomic orbital projected densities of states (PDOSs) of the TPyB-Co without SOC are illustrated in Figs. 3(d) and 3(e), respectively. As displayed in Fig. 3(d), six featured bands appear around the E$_{F}$, exhibiting a magnetic metallic states. These six featured bands primarily come from the $\emph{p}_{z}$ orbitals of C and N atoms of the TPyB molecule [Fig. 3(e)]. Figure 3(f) gives the calculated maximally localized Wannier functions (MLWFs) for the three spin-up bands around the E${_F}$ by using the Wannier90 package \cite{47}. Remarkably, the three optimized MLWFs exhibit a threefold rotational symmetry around the center of the TPyB molecule, having the same features as the $\emph{sp}^{2}$ hybridized orbitals. Hence, the TPyB molecule can be considered effectively as a super-atom and the six featured bands around the E$_{F}$ can be recognized to be formed by $\emph{sp}^{2}$ hybrid molecular orbitals, which can be equivalently transformed into $\emph{s}$, $\emph{p}_{x}$, and $\emph{p}_{y}$ molecular orbitals. The gaps opened in the $\emph{p}_{x}$ and $\emph{p}_{y}$ bands at the K/K$^{\prime}$ points are the result of the interactions from the $\emph{d}_{xy}$/$\emph{d}_{x^{2}-y^{2}}$ (and $\emph{d}_{z^{2}}$) orbitals at the B-site, as demonstrated in the TB model section. Although there is less overlap between the \emph{p} and \emph{d} orbitals in the energy range from -0.3 eV to 0.6 eV [Fig. 3(e)], strong interactions must exist between them, indicated by the obvious spin polarization of \emph{p} orbitals of the C and N atoms in the system. Since the A-site (TPyB molecule) owns $\emph{s}$, $\emph{p}_{x}$, and $\emph{p}_{y}$ molecular orbitals and the B-site (Co atom) contains $\emph{d}$ orbitals, the electronic and topological properties of the TPyB-Co can be physically studied by using the TB model Hamiltonian constructed above for the bitriangular lattice. Figure 4(a) shows the magnified four bands formed by $\emph{p}_{x}$ and $\emph{p}_{y}$ molecular orbitals (the solid curves), possessing the similar band features as given by the above TB model [Fig. 2(i)]. Note that in the above TB model for the bitriangular lattice, the $\emph{s}$ orbital is not considered for the A-site. Our TB model test calculations show that the s orbital does not vary the trends of the $\emph{p}_{x}$ and $\emph{p}_{y}$ orbitals at all if it is relatively far away from them. The traditional $\emph{s}-\emph{p}$ band inversion can not occur around the $\Gamma$ point unless the $\emph{s}$ orbital is close enough to the multiple $\emph{p}$ orbitals \cite{17}. The large band gaps of the $\emph{p}_{x}$ and $\emph{p}_{y}$ orbitals for both the spin-up and spin-down channels, opened at the K (or K$^{\prime}$) point in Fig. 4(a), can be well rationalized by the NN hopping between TPyB molecule and Co atom. Nevertheless, the band touching of the $\emph{p}_{x}$ and $\emph{p}_{y}$ orbitals at the $\Gamma$ point for each spin channel, with quadratic non-Dirac band dispersion, still exists, illustrated more clearly in Fig. 4(c).

When the SOC is turned on in the DFT calculations, a gap of 6.1 meV is opened around the E$_{F}$ ($\Delta_{1}$), as shown in Fig. 4(d), much larger than the gap opened in graphene \cite{50,51}. The reason is due to the strong SOC interactions here, enhanced by the $\emph{p}_{x}$ and $\emph{p}_{y}$ type molecular orbitals \cite{14} and the hybridization with Co atoms \cite{52}. It is also very commendable that the E$_{F}$ is exactly localized within the nontrivial gap in the material [Fig. 4(d)], which is favorable to experimental measurements of the electronic transports. The Berry curvature, Chern number, and edge states are calculated to identify the topological property of this gap ($\Delta_{1}$) by using the combination of DFT calculations and MLWF methods. Figure 5(a) shows the bands of TPyB-Co obtained from the Wannier interpolation approach (blue dots), matched perfectly with the DFT bands (solid black curves). The integration of the Berry curvatures over the first BZ [the inset of Fig. 5(a)] gives the Chern number of the system with $C=-1$, indicating the existence of the Chern insulating phase in the TPyB-Co. In addition, the anomalous Hall conductivity $\sigma_{xy}$ as a function of the energy is plotted in Fig. 5(b). A quantized Hall conductivity platform indeed appears around the E${_F}$, demonstrating an intrinsic Chern insulator of the TPyB-Co. The topologically nontrivial states can also be judged from the topological edge states of the target systems. By using the TB model constructed from the MLWFs, the band structure of the one-dimensional (1D) TPyB-Co nanoribbon is calculated. As displayed in Fig. 5(c), the blue curves denote the bulk states of the nanoribbon and the red curves denote the edge states. Figure 5(d) zooms in the bands in Fig. 5(c) around the E$_{F}$, from which the edge states are explicitly seen within the SOC gap, confirming the intrinsic Chern insulator of the system. The two edge states of A and B represent one chiral conducting channel appearing on each side of the sample, as illustrated by the schematic drawing in Fig. 5(e), consistent with the obtained Chern number of $C=-1$. The electronic structures of the TPyB-Co with the metastable FM state are also investigated (see Fig. S2 of the Supplemental Material [40]). The results reveal that the intrinsic QAH state also exists, indicating the robustness of the topological state in TPyB-Co material systems.

\section{\textbf{CONCLUSIONS}}

The band structure of the simple 2D triangular lattice in the basis of ($\emph{p}_{x}$, $\emph{p}_{y}$) contains band degeneracy not only at the $\Gamma$ point but also at the K/K$^{\prime}$ points, with quadratic non-Dirac and linear Dirac band dispersions, respectively. The topological properties of the system are simultaneously determined by the non-Dirac and Dirac bands, quite different from the previous widely studied honeycomb lattice. Based on TB-model Hamiltonian calculations, the global SOC-induced gap in the system is found to be topologically trivial rather than nontrivial due to the interesting destructive interference effect of the topological states at the $\Gamma$ and K/K$^{\prime}$ points. To release the topological behaviors of the system, a TB model Hamiltonian of a bitriangular lattice is constructed, which is formed by two triangular lattices [with ($\emph{p}_{x}$, $\emph{p}_{y}$) and ($\emph{d}_{xy}$, $\emph{d}_{x^{2}-y^{2}}$) orbitals, respectively]. The topologically nontrivial states can be achieved in this bitriangular lattice due to the breakdown of the destructive interference effect, inferring that the bitriangular lattice can be a genre of a very important prototype model for exploring the QSH state or the QAH state beyond the widely studied honeycomb lattices. The orbitals on the other triangular lattice in the bitriangular lattice model can also be extended to other orbitals, such as ($\emph{p}_{x}$, $\emph{p}_{y}$) and $\emph{d}_{z^{2}}$. Our DFT calculations predict that this type of 2D triangular lattice metal-organic framework TPyB-Co is an intrinsic Chern insulator, rationalized well by the constructed bitriangular TB model. The nearest-neighbor hopping between the TPyB molecule and Co atom removes the topological behaviors at the K and K$^{\prime}$ points. The coupling effect of the topological states is thus eliminated, leading to the release of the topological state at the $\Gamma$ point. The proposed TB model Hamiltonian will provide theoretical guidance for the search for and design of 2D topological quantum states in a new class of material systems with triangular lattices. Our findings may greatly push the experimental realization of QSH and QAH states in realistic 2D material systems.

\begin{acknowledgments}
This work was supported by the National Natural Science Foundation of China under Grants No. 11574051, No. 11604134, and No. 11747137, the Natural Science Foundation of Shanghai under Grant No. 14ZR1403400, the Natural Science Foundation of Jiangsu Province (China) under Grant No. BK20170376, and the Natural Science Foundation of Higher Education Institutions of Jiangsu Province (China) under Grant No. 17KJB140023.

J. Y. Zhang and B. Zhao contributed equally to this work.
\end{acknowledgments}


\end{document}